\documentclass[utf8]{frontiersSCNS_arxiv} 

\usepackage{url,hyperref,lineno,microtype,subcaption}
\usepackage[onehalfspacing]{setspace}
\usepackage[final]{changes}
\definecolor{darkgreen}{RGB}{0,100,0}
\makeatletter
\@namedef{Changes@AuthorColor}{darkgreen}
\colorlet{Changes@Color}{darkgreen}

\def\keyFont{\fontsize{10}{11}\helveticabold }
\def\firstAuthorLast{Mattsson {et~al.}} 
\def\Authors{Carolina Mattsson $^{1,7,*}$, Frank W. Takes $^{1,8}$, Eelke M. Heemskerk $^{2,8}$, Cees Diks $^{3,9}$, Gert Buiten $^{4}$, Albert Faber $^{5}$, Peter M.A. Sloot $^{6,10,11,12,13}$}
\def\Address{$^{1,}$ Computational Network Science Lab, Leiden Institute of Advanced Computer Science, Leiden University, Leiden, Netherlands \\
$^{2}$ Department of Political Science, University of Amsterdam, Amsterdam, Netherlands\\
$^{3}$ CeNDEF, Faculty Economics and Business, University of Amsterdam, Amsterdam, Netherlands \\
$^{4}$ Statistics Netherlands, Den Haag, Netherlands \\
$^{5}$ Ministry of Economic Affairs \& Climate, Den Haag, Netherlands \\
$^{6}$ Computational Science Lab, Faculty of Science, University of Amsterdam, Amsterdam, Netherlands \\
$^{7}$ Network Science Institute, Northeastern University, Boston, MA, USA \\
$^{8}$ CORPNET, University of Amsterdam, Amsterdam, Netherlands \\
$^{9}$ Tinbergen Institute, Amsterdam, Netherlands \\
$^{10}$ Institute for Advanced Study, University of Amsterdam, Amsterdam, Netherlands \\
$^{11}$ Complexity Institute, Nanyang Technological University, Nanyang, Singapore \\
$^{12}$ Complexity Science Hub Vienna, Vienna, Austria \\
$^{13}$ ITMO University, Saint Petersburg, Russia}

\def\corrAuthor{Leiden Institute of Advanced Computer Science \\ Niels Bohrweg 1, 2333CA Leiden, Netherlands}
\def\corrEmail{e.s.c.mattsson@liacs.leidenuniv.nl\\mattsson.c@northeastern.edu}

\begin{document}
\onecolumn
\firstpage{1}

\title[Functional structure]{Functional structure in production networks}

\author[\firstAuthorLast ]{\Authors} 
\address{} 
\correspondance{} 

\extraAuth{}

\maketitle

\begin{abstract}

Production networks are integral to economic dynamics, yet dis-aggregated network data on inter-firm trade is rarely collected and often proprietary. Here we \replaced{situate}{analyze} company-level production networks \replaced{among networks from other domains according to their}{the} local connectivity structure. \added{Through this lens, we study a regional and a national network of inferred trade relationships reconstructed from Dutch national economic statistics and re-interpret prior empirical findings.} We find that company-level production networks have so-called functional structure, as previously identified in protein-protein interaction (PPI) networks. Functional networks are distinctive in their over-representation of closed squares, which we quantify using an existing measure called spectral bipartivity. Shared local \added{connectivity} structure \replaced{}{of production and PPI networks} lets us ferry insights between domains. PPI networks are shaped by complementarity, rather than homophily, and we use multi-layer directed configuration models to show that this principle explains the emergence of functional structure in production networks. Companies are especially similar to their close competitors, not to their trading partners. \replaced{Our findings have practical implications for the analysis of production networks and a thorough understanding of their local connectivity structure will help us better reason about the micro-economic mechanisms behind their routine function, failure, and growth.}{Overall, that production networks have functional structure at the resolution of companies suggests methods from network biology and network ecology may be beneficial for uncovering microeconomic mechanisms behind economic dynamics.}

\tiny
 \keyFont{ \section{Keywords:} production networks, inter-firm networks, complextiy economics, official statistics, trade linkages, functional networks, bipartivity} 
\end{abstract}

\section{Introduction}

It has become established knowledge within complexity economics~\citep{arthur_foundations_2021} that network structure affects economic dynamics over the short-, medium-, and long- term. Cascading processes over networks have been used to model supply disruptions that propagate in a matter of days or weeks~\citep{burkholz_systemic_2016,inoue_firm-level_2019}. Trade linkages have been used to explain aggregate fluctuations in business activity that play out over months or years~\citep{acemoglu_network_2012,carvalho_production_2018}. Structural changes happen over decades and we know that national growth trajectories are strongly affected by the network structure of economic activity~\citep{hausmann_how_2013,mcnerney_how_2018}. The routine function, failure, and growth of economic production networks are at the heart of these dynamics.

\added{The mechanisms underlying dynamics on and of production networks are thought to operate at the level of} trade relationships among individual companies~\citep[see:][]{hazama_measuring_2017,carvalho_production_2018,inoue_firm-level_2019}. However, \replaced{}{there are compelling reasons to turn our focus towards networks of trade relationships among individual companies. Such dis-aggregated data on production networks facilitates the incorporation of microeconomic analysis~\citep[see:][]{carvalho_production_2018} and allows more realistic simulation of economic systems~\citep{hazama_measuring_2017,inoue_firm-level_2019}.} \added{our understanding of company-level production networks is limited;} empirical network data on customer-supplier ties is not often available and, when it is, generally proprietary~\citep{fujiwara_large-scale_2010,ohnishi_network_2010,magerman_belgian_2015}. \added{Moreover, findings based on such data remain difficult to interpret in the context of wider research on production networks because they combine local and national scales. Trade relationships are more often considered \emph{either} in detailed, local case studies \emph{or} as} aggregated trade linkages among sectors, industries, and countries based in officially-prepared macroeconomic statistics~\citep{uzzi_social_1997,coenen_local_2010,acemoglu_network_2012,miller_output_2017,mcnerney_how_2018}. 

\added{In this work, we situate company-level production networks within a wider space of networks that are different in nature, but similar in structure. For this we develop a typology based on well-known and recognizable \emph{local connectivity structures}.} Random networks, social networks, and two-mode networks are the most important and well-known connectivity types \added{and they form the basis for our typology}~\citep{newman_random_2001,rivera_dynamics_2010, borgatti_network_1997}. Commonalities among networks with the same local connectivity type present opportunities to use established knowledge from one domain to better understand networks in another. Specifically, a new network type with a \added{distinctive local connectivity structure} has recently been identified in work on protein-protein interaction (PPI) networks~\citep{kovacs_network-based_2019,kitsak_latent_2020}. \added{This adds so-called functional networks to our typology. We highlight several existing empirical findings~\citep{ohnishi_network_2010,fujiwara_large-scale_2010} to suggest that company-level production networks are best characterized as functional networks and go on to explore this hypothesis.}

Specifically, we consider a regional and a national company-level production network reconstructed from Dutch national economic statistics. \added{As detailed in~\citet{hooijmaaijers_methodology_2019}, Statistics Netherlands (CBS) has used official statistics to} systematically infer customer-supplier ties among companies in the Netherlands for each of $677$ product groups (e.g., ``Electricity'', ``Fertilizer'', ``Shipping services'', etc.). Starting from this multi-layer network, we analyze the local connectivity structure of the network of unique (inferred) trade relationships among companies with $5+$ employees within Zeeland province and the whole of the Netherlands. \added{We then generalize the CBS reconstruction process to produce ensembles of networks of (hypothetical) trade relationships among the Zeeland companies with $5+$ employees; this lets us study how local connectivity structure emerges in} company-level production networks. 

To asses\added{s} whether networks have functional structure, we re-purpose an existing measure that captures a distinctive feature of local connectivity in functional networks. Link prediction performance on PPI networks suggests that three-step closure is more common than two-step closure~\citep{kovacs_network-based_2019}. That is, closed squares are more prominent than are closed triangles. We measure this using spectral bipartivity, which quantifies the abundance of even versus odd cycles in a network’s local connectivity structure~\citep{estrada_network_2016}, in comparison to random expectation; functional networks have higher-than-random spectral bipartivity.

We do indeed find functional structure in both the national and the regional company-level production networks. Compared to randomized versions of itself, the \added{reconstructed} regional network has significantly higher spectral bipartivity (Kolmogorov-Smirnov (KS) statistic $1.0$, $N_1 = 1$, $N_2 = 1000$, $p < 0.001$). The value itself remains much smaller than~$1$, indicating that the network structure is definitely not bipartite as would be the case if it were a two-mode network. Small values then let us approximate the logit-transformed spectral bipartivity and extend our results to the larger Netherlands network. This measure is again higher than expected ($KS = 1.0$, $N_1 = 1$, $N_2 = 25$, $p < 0.04$) indicating functional structure.

In the literature on protein-protein interactions, network structure is thought to be shaped by the principle of \textit{complementarity} as opposed to the principle of \textit{homophily} well known to shape social networks~\citep{kitsak_latent_2020,mcpherson_birds_2001}. Complementarity in PPI networks reflects the practical fact that proteins physically bind with one another at compatible binding sites~\citep{kovacs_network-based_2019}. \replaced{This is a useful concept for company-level production networks, as well, since trade relationships imply the exchange of some product between companies that hold complementary roles as customer and supplier.}{one company supplies some product that another uses}. Trade compatibility imposes a specific constraint on the formation of trade relationships that might explain the emergence of functional structure in production networks.

To test this explanation, we use multi-layer directed configuration models to generate ensembles of networks made up of trade-compatible relationships within Zeeland. \added{In each layer, we make a random selection of the possible ties between customers and suppliers of products in a product group}. By defining trade compatibility according to progressively more detailed product categorizations, we ramp up the strictness of the \replaced{}{complementarity} constraint. These generated networks are then compared against their randomized version where customer-supplier complementarity has been broken. Our analysis finds \replaced{that imposing more stringent complementarity}{more detailed product categorizations} introduces consistently and significantly more functional structure ($KS = 1.0$ , $N_1 = 25$, $N_2 = 25$, $p < 10^{-14}$).

Identifying functional structure in company-level production networks has practical implications for the analysis of such networks and wider implications for network science and complexity economics. Under the logic of functional networks, companies trade with complementary others and are especially similar to their close competitors, not their trading partners. This also implies that production networks are markedly different from social and economic networks driven by homophily; the most comparable networks are PPI networks and food webs. More generally, we have demonstrated the usefulness of network categorization according to a typology of local connectivity structure. Going forward, \replaced{a thorough understanding of the local connectivity structure of}{having identified functional structure in} production networks will help us better reason about the micro-economic mechanisms behind their routine function, failure, and growth.

The remainder of the paper is structured as follows. In Section~\ref{sec:theory}, we describe several types of networks with distinct local connectivity structure \added{and discuss in detail highly relevant prior work}. Section~\ref{sec:datamethod} describes the company-level production networks that we study and the methods we use to characterize their local connectivity structure. In Section~\ref{sec:results} we present our findings and in Section~\ref{sec:discussion} we discuss their implications.

\section{Theory} \label{sec:theory}

\added{Networks studied in different domains can nonetheless be similar in structure. Section~\ref{sec:typology} develops a network typology based around the local connectivity structures of random, social, and two-mode networks. To this typology we add recently identified so-called ``functional'' networks. In Section~\ref{sec:literature} we advance the hypothesis that company-level production networks are best characterized as functional networks.}

\subsection{Network typology} \label{sec:typology}

Networks can be categorized into typologies wherein different network types have systematically different structural properties. Many network analysis tools have been developed especially for use with networks of some particular type~\citep{opsahl_triadic_2013,masuda_clustering_2018}. The same algorithms can have markedly different performance across network types~\citep{ghasemian_stacking_2020} and the same measures will often have different typical ranges~\citep{newman_structure_2003,costa_characterization_2007}. Network measures also tend to be correlated~\citep{jamakovic_relationships_2008,bounova_overview_2012} with local network features affecting global ones~\citep{colomer-de-simon_deciphering_2013,jamakovic_how_2015,asikainen_cumulative_2020}. As such, we use the following network typology based in recognizably distinct \emph{local connectivity structures}. 

\emph{\helvetica Random networks} \hspace{6pt}
Random networks are those where the set of existing links might have come to occur by chance. Several celebrated network properties, such as the emergence of a \emph{giant component} and the \emph{small world} property, are identifiable already in random networks~\citep{bollobas_random_2001,newman_random_2001}. Other common network properties, such as the existence of \emph{hubs} or \emph{communities}, can be introduced using block-wise random networks~\citep{holland_stochastic_1983,newman_random_2001,karrer_stochastic_2011}. This flexibility makes random networks especially useful as a baseline comparison for empirical network data and generative network models~\citep{costa_characterization_2007}. Some measures, such as modularity~\citep{newman_modularity_2006} and degree assortativity~\citep{pastor-satorras_dynamical_2001,newman_mixing_2003}, directly incorporate a comparison to random expectation.

\emph{\helvetica Two-mode networks} \hspace{6pt}
Two-mode networks are those where the links are affiliations between categorically different nodes. For instance, directors are affiliated with the corporate boards on which they serve~\citep{mizruchi_what_1996,seierstad_for_2011,takes_centrality_2016,valeeva2020duality}. Two-mode networks are \emph{bipartite} in that the nodes can be separated into two groups where links exist between, but not within, groups~\citep{borgatti_network_1997,holme_network_2003}. This is a hard constraint on the existence of network links that affects the interpretation of any network-structural property. As such, there are specific network analysis techniques and particular versions of centrality, modularity, closure, and other measures developed for two-mode networks~\citep{borgatti_network_1997,zhou_bipartite_2007,barber_modularity_2007,opsahl_triadic_2013,berardo_bridging_2014}.

\emph{\helvetica Social networks} \hspace{6pt}
Social networks are those where links exist between nodes who associate with one another in some sense. This is a soft constraint on the formation of links that affects many network properties. Social networks have high \emph{clustering} as various social dynamics are known to generate \emph{triadic closure}~\citep[see:][]{rivera_dynamics_2010}. They are typically shaped by \emph{homophily} leading to high \emph{assortativity}, where nodes who associate with one another tend to be similar in some sense~\citep{mcpherson_birds_2001,newman_mixing_2003}. Social networks frequently contain interpretable \emph{communities}~\citep{davis_deep_2009,blondel_fast_2008} and are often assortative in degree~\citep{johnson_entropic_2010}. It has been suggested that the tendency of nodes to form communities may be a derivative feature of triadic closure operating locally~\citep{jamakovic_how_2015,colomer-de-simon_deciphering_2013}. The over-representation of triangles also allows link prediction algorithms that operate on a two-step basis (L2) to perform well on social networks~\citep{ghasemian_stacking_2020}. Degree assortativity may also be related to triadic closure, homoplily, and community structure~\citep{newman_why_2003,asikainen_cumulative_2020}.

\emph{\helvetica Functional networks} \hspace{6pt}
Functional networks are those where links form between nodes that complement one another in fulfilling some function. This network type has been identified in work on protein-protein interaction (PPI) networks; proteins bind not with similar others but with those that have a binding site physically compatible with their own~\citep{kovacs_network-based_2019}. These networks are shaped by the principle of \textit{complementarity}, a different soft constraint that likely also affects many, related, network properties~\citep{kitsak_latent_2020}. \citet{kovacs_network-based_2019} establishes that L2 heuristics under-perform in link prediction on PPI networks while those operating on a three-step basis (L3) are more accurate. In analogy with social networks, this might suggest that functional dynamics generate \emph{tetradic closure}. Meso-scale structures in PPI networks have been termed \emph{functional modules} as they are often interpretable as key to some higher-level cellular function~\citep{barabasi_network_2004,chen_detecting_2006,ghiassian_disease_2015}. \emph{Disassortativity} in degree may be an expected property of functional networks more broadly~\citep{barabasi_chapter_2016,johnson_entropic_2010}. Food webs and interdisciplinary collaboration networks have been proposed as additional domains where the same structural patterns are likely to be found~\citep{kitsak_latent_2020}. 

Table~\ref{tab:example} highlights the difference in local connectivity structure between social and functional network types, with links likely to form shown using dashed lines. In social networks links form between nodes who associate with one another, nodes who associate with one another are similar in some sense, and various social dynamics generate closed triangles. Social networks thus have a higher clustering coefficient than random networks. The analogous intuition for functional networks is that links form between nodes in fulfilling their function, that nodes who complement the same partners are similar in some sense, and that various functional dynamics generate closed squares. Functional networks would score higher than random networks on measures that capture this aspect of local connectivity structure.

\begin{table}[!tb]
    \caption{Key differences between social and functional networks.\\}
    \label{tab:example}
    \scriptsize
    \begin{tabular}{l  m{0.15\textwidth} m{0.25\textwidth} m{0.15\textwidth} l l l } 
         & \textbf{Toy network} & \textbf{Example}
         & \textbf{Principle} & \textbf{Closure} & \textbf{Meso-scale} \\
     \textbf{Social} & \includegraphics[width=0.15\textwidth]{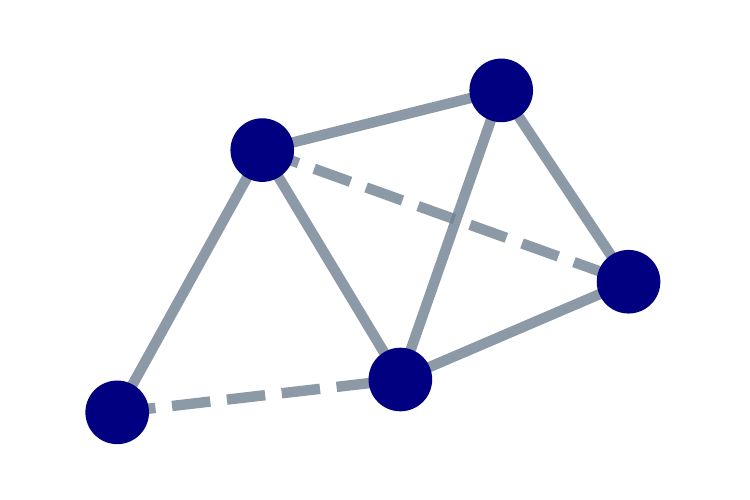} & People are often friends with others similar to themselves, and a pair of friends likely also have friends in common. & Homophily & Triadic & Community  \\
     \textbf{Functional} & \includegraphics[width=0.15\textwidth]{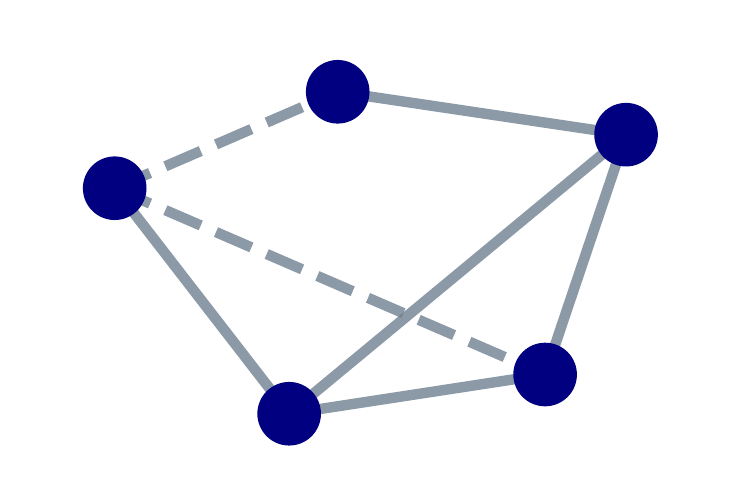} & Proteins interact at a physical binding site; proteins are likely to bind if one is similar to the other's other partners. & Complementarity & Tetradic & Module   \\
    \end{tabular}
\end{table}

\subsection{ \added{Features of company-level production networks} } \label{sec:literature}

\added{
Characterizing local connectivity structure directly, as we do here, offers a new lens through which to consider known structural features of networks. Findings within existing studies of empirical company-level production networks are thus exceedingly relevant. Here we discuss in detail two papers that describe the structure of a Japanese inter-firm network produced by Tokyo Shoko Research Ltd~\citep{ohnishi_network_2010,fujiwara_large-scale_2010}. This is a company-level production network where nodes are Japanese firms and links correspond to trade relationships in materials or services; financial relationships are downplayed. Several findings in these papers let us form the hypothesis that company-level production networks have functional structure.

\citet{ohnishi_network_2010} perform a comprehensive analysis of three-node motifs on the Japanese inter-firm network from 2005. On a simple, directed network there are thirteen possible three-node motifs and this paper presents their empirical prevalence compared to random expectation~\citep[][Fig.~4~\&~5]{ohnishi_network_2010}. Recall that closed squares, but not closed triangles, feature prominently in the local connectivity structure of functional networks. In this paper, the most substantially over-represented motifs in the Japanese inter-firm network are two-link, V-shaped motifs; these are the most compatible with closed squares. Three-link, feedforward and feedback loops forming closed triangles are found to be the most substantially under-represented. These findings can be interpreted as empirical evidence of functional structure in this company-level production network.

\citet{fujiwara_large-scale_2010} conduct a multi-pronged network analysis of the Japanese inter-firm network compiled in September 2006. Three of their findings are especially relevant with respect to our typology of local connectivity structure. First, this paper shows the network exhibits disassortativity in degree~\citep[][Fig.~3]{fujiwara_large-scale_2010}. Recall that disassortativity may be an expected feature of functional networks as is assortativity for social networks. Second, this paper performs a detailed analysis of meso-scale structure within the sub-network defined by firms in the manufacturing sector. Using modularity maximization they identify a large number of small groups with a striking qualitative interpretation: ``From the database of the information on the firms, we found that many of those small communities are each located in same geographical areas forming specialized production flows. An example is a small group of flour-maker, noodle-foods producers, bakeries, and packing/labeling companies in a rural area''~\citep[][pg.~570]{fujiwara_large-scale_2010}. While these groups are referred to as ``communities'', they might be better understood as so-called ``functional modules'' in that they are key to the functioning of economic production at a higher level. Finally, this paper finds locally bipartite structure within prominent (sub-)industries also identified by modularity maximization. In each (sub-)industry there is a handful of large, recognizable firms who are not often directly linked, as they are competitors, but have many suppliers (and customers) in common~\citep[][pg.~570]{fujiwara_large-scale_2010}. Locally bipartite meso-scale structure means there are many closed squares, enough that these groups are picked up by modularity maximization. Closed squares, functional modules, and disassortativity are all hallmarks of functional networks. 
}

\section{Data and Methods} \label{sec:datamethod}

This paper makes use of a network dataset produced by Statistics Netherlands (CBS) to explore our hypothesis that company-level production networks have functional structure. In Section~\ref{sec:data} we describe the data itself and how we make use of it. Section~\ref{sec:method} presents our methods, describing how we characterize local connectivity structure.

\subsection{Data} \label{sec:data}

CBS has produced a network dataset of systematically inferred customer-supplier ties among Dutch companies~\citep{hooijmaaijers_methodology_2019}. In Section~\ref{sec:networkconstruction} we \added{describe how we define a national and a regional company-level production network from this data; we also briefly} relay how this data was produced. Section~\ref{sec:ensemble} details how we generalize the CBS reconstruction process, in this work, to construct ensembles of many possible customer-supplier networks within Zeeland province.

\subsubsection{National and regional company-level production networks}
\label{sec:networkconstruction}

We \added{begin with an existing dataset of inferred customer-supplier ties} produced by researchers at CBS~\citep{hooijmaaijers_methodology_2019}. \replaced{}{detail this approach: used} \added{This network has} $677$ layers\added{, each} correspond\added{ing} to \added{domestic} trade in a particular product group.\replaced{}{, where links are directed from supplying to using companies} Product groups include ``Electricity'', ``Fertilizer'', ``Shipping services'', and ``Accounting \& tax administration'' as in the Dutch implementation of the European Classification of Products by Activity~\citep[][CPA 2008]{eurostat_nace_2008}. Companies likely to \emph{supply} products from a product group \replaced{}{traded in the domestic market} \added{were matched} with companies likely to \emph{use} such products, with inferences based in Dutch economic statistics for the year 2012. \added{Customer-supplier ties in each layer should be understood to reflect a selection of trading pairs among sets of likely customers and suppliers for products in that product group.} Table~\ref{tab:descriptives} describes the reconstructed network of inferred customer-supplier ties within the Netherlands and within Zeeland province; self-loops have been removed. Zeeland is a low-lying delta region home to 381,407 people in 2012~\citep{cbs_bevolkingsontwikkeling_2012}. 

\begin{table}[!tb]
    \caption{Description of \replaced{the data provided by CBS. Highlighted are the reconstructed networks for}{our production networks of} the Netherlands and Zeeland province \replaced{above our}{for different} company size threshold (measured using the number of employees). Also shown are the statistics for the giant connected component (GCC) of these networks.}
    \label{tab:descriptives}
    \scriptsize
    \vspace{0.6em}
    \begin{tabular}{r|rrrrrrr}
     &   &  &   & \replaced{}{Unique} & \textbf{\added{GCC}} & \textbf{\added{GCC}} & \textbf{\added{GCC}} \\
     & \textbf{Nodes}  & \textbf{Layers} & \textbf{Multi-edges}  & \textbf{\added{Simple} edges} & \textbf{\added{Nodes}} & \textbf{\added{Multi-edges}} & \textbf{\added{Simple edges}} \\
    \hline
    \textbf{Netherlands:} & & & & &&\\ 
    \added{All}      &  875,222 & 677   & 310,324,477 & 195,903,806 & 875,222 & 310,324,477 & 195,903,806 \\
    5+       &  102,461 & 677   & 116,652,466 &  50,930,077 & 102,461 & 116,652,466 &  50,930,077 \\
    \textbf{Zeeland:}    & & & & &&\\ 
    \added{All}      &   18,398 & 667   &   3,500,797 &   2,143,412 &  18,337 &   3,500,762 &   2,143,379 \\
    5+       &    2,497 & 652   &     848,015 &     334,334 &   2,497 &     848,015 &     334,334 \\
    \end{tabular}
\end{table}

\begin{figure}[h!] 
\begin{center}
\includegraphics[height=5.25cm]{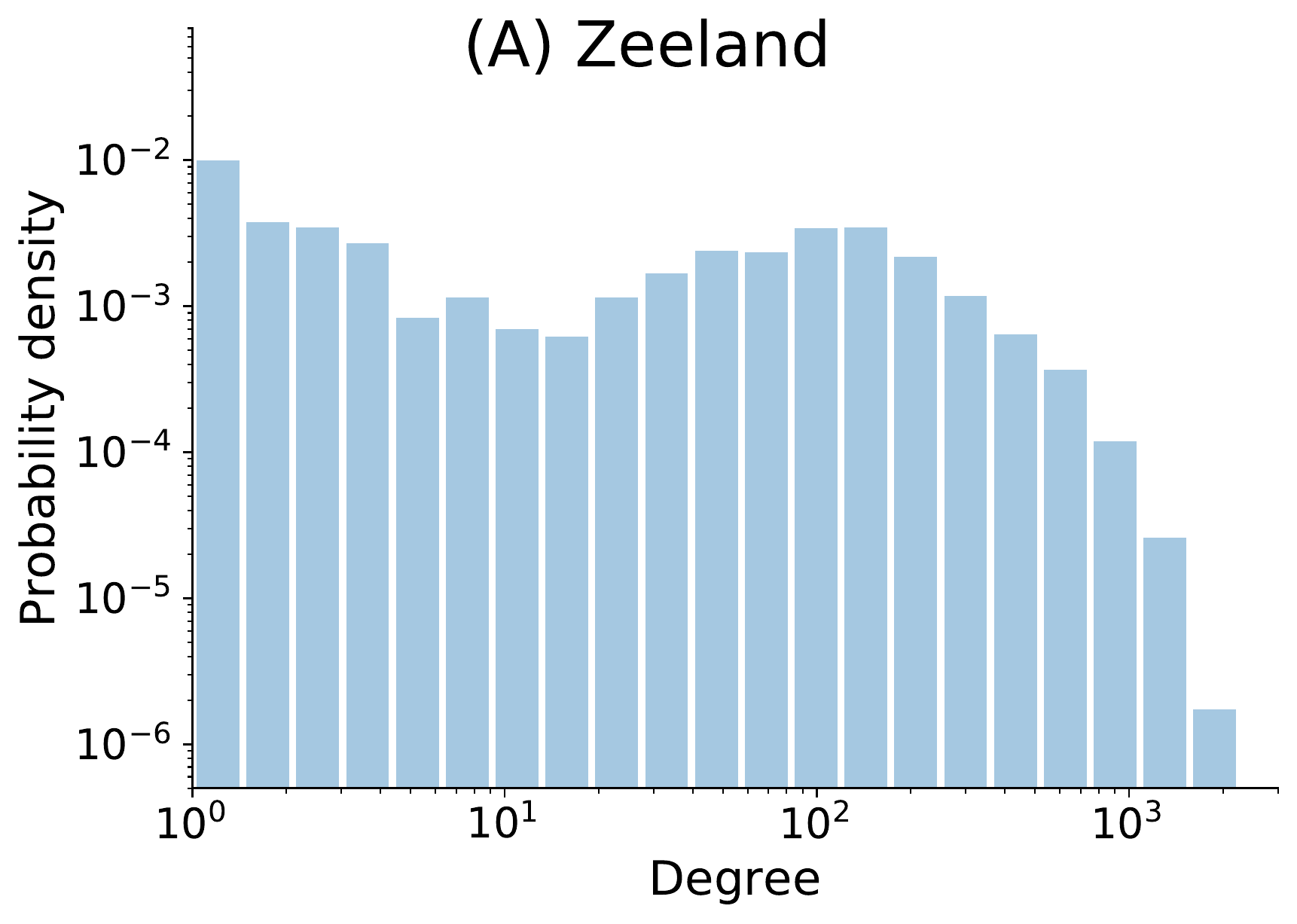}
\includegraphics[height=5.25cm]{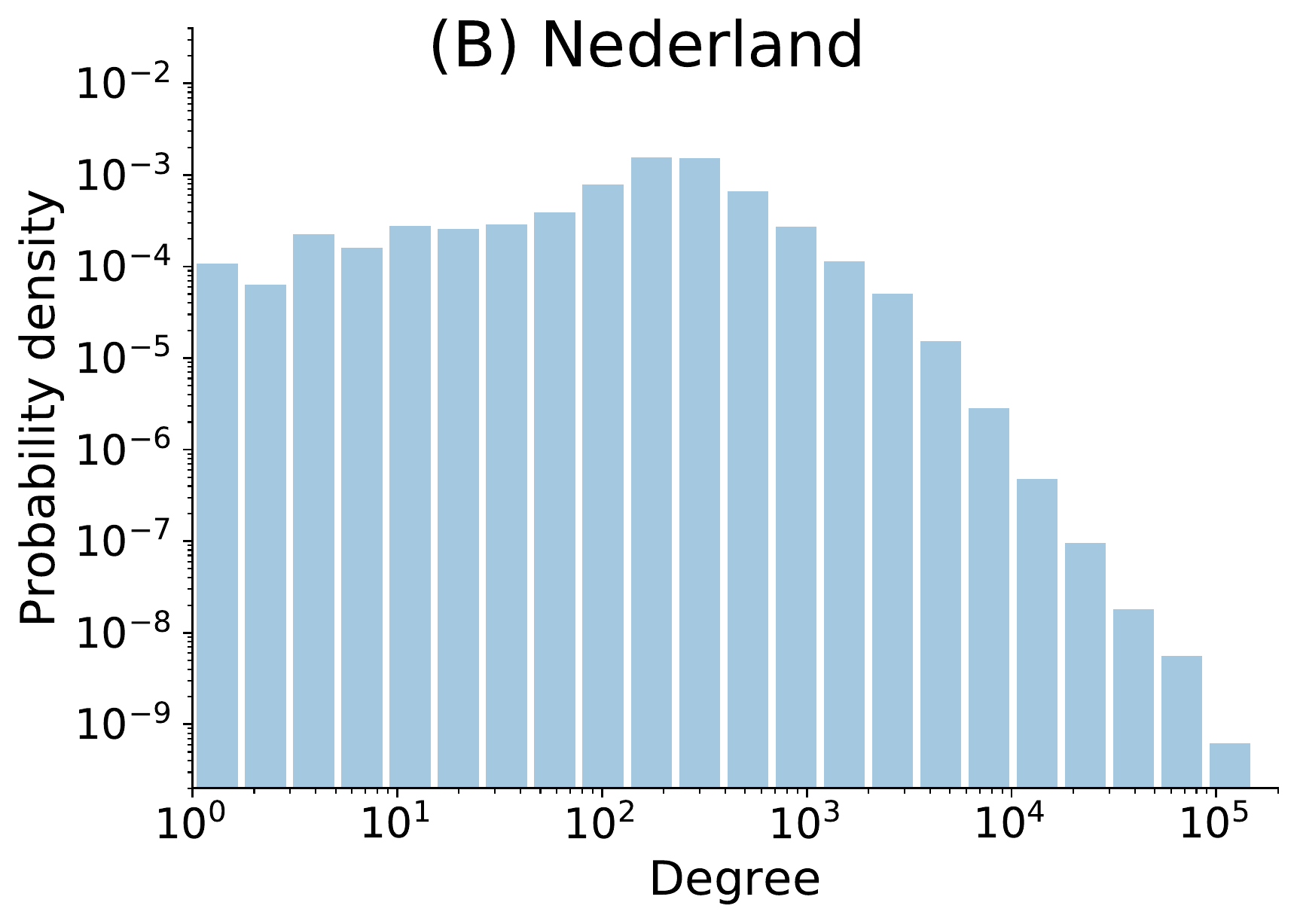}
\end{center}
\caption{\added{Degree distributions of (A) Zeeland and (B) Netherlands production networks reflecting the unique, unweighted, undirected (inferred) trade relationships among companies with $5+$ employees.}}\label{fig:degrees}
\end{figure}

Based on the CBS data, this work defines a national and a regional company-level production network. \added{For this,} we first consider the \added{sub-}network among companies with five or more employees \added{within Zeeland province and the whole of the Netherlands.} We then focus on the ``simple'' versions of these multi-layer, directed, customer-supplier networks. Simple networks are those with unique, unweighted, undirected links. That is, we place an (inferred) trade relationship between any two companies with an inferred customer-supplier tie, in either direction, in any of the product groups. To give an idea of the heterogeneity in \added{our version of these} networks, Figure~\ref{fig:degrees} describes \replaced{their degree distributions on a log-log scale}{also the sub-networks among companies with at least $20$, $100$, and $250$ employees. For completeness, we include also the range of the number of edges found across $25$ randomizations as defined in Section~\ref{sec:randomization}.} \added{Filtering out companies with few employees is done} as the underlying, company-level statistics are deemed insufficiently reliable for very small companies; these are primarily low-degree nodes. Zeeland province is deemed a suitable sub-network on which to focus for its small size and its somewhat lesser integration, geographic and economic, with the rest of the Netherlands. It is also considered an industrial cluster by policy makers and so is expected to have some general internal coherence.

\added{\emph{\helvetica CBS network reconstruction process} \hspace{6pt}} Dutch national Supply- and Use- Tables \replaced{include aggregated}{are available from} trade flows \added{between industries} by product group~\citep{eurostat_eurostat_2008}. \added{These are official statistics and extensively validated. CBS has applied systematic inference to} dis-aggregate \added{the domestic portion of} these flows down to the company level\added{. Customer-supplier ties are inferred based on assumptions made about individual companies with regard to their size, their location, and their role with respect to each product group. Used in this are economic statistics} collected via official surveys that include total turnover, geographic location, and industry code according to the Dutch implementation of the Statistical classification of economic activities in the European Community~\citep[][NACE Rev.~2]{eurostat_nace_2008}. \added{These surveys are mandatory and use a stratified sampling framework on the number of employees. Note that their aim is to achieve statistically representative, not absolute, accuracy and so inferences are noisy especially for smaller companies. Each product group in the Dutch CPA (2008) becomes a layer in the reconstructed network. Within a layer,} companies \added{active in an industry that uses} products classified into \replaced{that}{a} product group are assigned a number of suppliers\replaced{, i.e. an}{ captured by its} in-degree. \added{It was assumed that most companies deal with few suppliers per product-group;} in-degree is kept small. Out-degree, analogously, represents the number of customers for a company \added{active in an industry that supplies} products in that product group. \added{Estimations of out-degree were made to reflect} empirical distributions with respect to company size \replaced{known from}{,in analogue to} studies done in Japan~\citep{watanabe_relations_2013}. \added{Also meant to be representative, on average, out-degree is noisy for individual companies.} \replaced{Finally,}{Precisely which} suppliers are matched with customers. \added{Matching} was done using limited empirical data on known trade relationships combined with information on geographic distance. The assumption that geographic distance matters is well-supported~\citep{dhyne_three_2016,bernard_production_2019,carrere_gravity_2020}. \added{For further details of the CBS network reconstruction process please consult \citet{hooijmaaijers_methodology_2019}.}

\subsubsection{Multi-layer complementary configuration models} \label{sec:ensemble}

The network theory laid out in Section~\ref{sec:theory} suggests a specific \replaced{reason}{way} functional structure might arise in production networks: trade connects companies with complementary roles\added{, i.e. customer and supplier,} in the \replaced{trade of}{production and/or distribution} of particular goods. Simulated data is used to asses\added{s} this explanation. The CBS reconstruction process is already based in matching customers and suppliers where the industry pairing suggests the two companies are compatible in trade, and here we generalize this process in two ways. First, instead of one reconstructed network for Zeeland we consider an ensemble of many possible networks among Zeeland firms with $5+$ employees. \added{Second, we leverage} hierarchical structure of the European CPA (2008). The coarser levels of this product classification are standardized across Europe; commonly studied are the top-level sector products (CPA Level 1) and progressively more detailed CPA Levels 2, 4, and 6. The Dutch national implementation (here, CPA National) is a minor refinement of CPA Level 6.

To generate ensembles of hypothetical company-level production networks we employ a multi-layer directed configuration model. The configuration model is a standard approach for generating ensembles of networks by randomly pairing edge stubs~\citep[][v2.5]{newman_random_2001,hagberg_exploring_2008}. We retain the in- and out-degree sequences from \replaced{}{the} \added{each product-layer of the CBS network} reconstruction \added{among Zeeland companies with $5+$ employees, without self-loops,} while drawing many possible wiring instances from the multi-layer configuration model. Our implementation generates a random, directed, multi-network \emph{per product group} in the classification; it then combines the layers into one customer-supplier network. At each CPA Level, the process is repeated multiple times to generate multiple independent realizations from the multi-layer configuration model. The result is a set of networks where ties are trade-compatible relationships under \replaced{the}{a complementarity} constraint \added{that trade occurs between companies supplying a product and companies using that same product}.  \added{Customer-supplier complementarity} grows progressively more stringent as we use more detailed product categorizations to define these roles.

Wherever the directed configuration model introduces self-loops, a known artefact of stub-matching~\citep[see:][]{newman_random_2010}, we rewire away the offending links. Specifically, we pair each self-loop with a random other edge and swap their target node~\citep[see:][Figure 1(a)]{hanhijarvi_randomization_2009}. On the other hand, when the configuration model generates multi-edges, i.e. edges between the same two nodes, these are allowed to remain; the networks are already multi-layer with many multi-edges.

Table~\ref{tab:model} describes the complementary ensembles, giving the network statistics found across $25$ independent realizations at each of the highlighted levels of the European CPA (2008). The resulting multi-layer networks maintain the total number of customer-supplier ties as well as the degree and role of each company in supplying and/or using products with a particular classification. \added{The hierarchical CPA levels} introduce variation in the number of product groups considered separately, i.e. the number of layers. \added{The complementarity constraint---}that customer-supplier ties may only go from companies who supply to those who use products \added{represented in that layer---is stronger at finer levels}. Notice the relative similarity of CPA Level 6 and CPA National. \added{As before, we focus in this work on the simple version of these networks where links correspond to unique trade-compatible relationships.} \added{The configuration model} introduces randomness in \added{the wiring diagram of the networks and} the number of unique trading relationships.

\begin{table}[!tb]
    \caption{Network statistics for the multi-layer complementary ensembles over $25$ realizations, \added{simulating trade-compatible relationships among companies with $5+$ employees within Zeeland province. Layers are designated according to the hierarchical levels of the European CPA (2008) and its Dutch implementation.}}
    \vspace{0.6em}
    \label{tab:model}
    \scriptsize
    \begin{tabular}{l|rrrrrrrr}
         & \textbf{Nodes}  & \textbf{Multi-edges}  & \textbf{Layers} &  \multicolumn{2}{l}{\textbf{Simple edges}} \\
    \hline
     &    &   &  &        min &    max   \\
    \multicolumn{1}{r|}{\textbf{Zeeland, 5+:}} & & & & \\ 
    CPA Level 1  &  2497 & 848,015 &  18 & 371,446 & 372,405 \\
    CPA Level 2  &  2497 & 848,015 &  79 & 369,483 & 370,190 \\
    CPA Level 4  &  2497 & 848,015 & 391 & 369,303 & 369,977 \\
    CPA Level 6  &  2497 & 848,015 & 614 & 368,324 & 368,978 \\
    CPA National &  2497 & 848,015 & 652 & 368,007 & 368,813 \\
    \end{tabular}
\end{table}

\subsection{Methods} \label{sec:method}

In this study we analyze the local connectivity structure of the production networks for Zeeland, the whole of the Netherlands, and our \replaced{complementary}{multi-layer} ensemble. \replaced{}{The simple versions of these networks are those where ties represent unique, undirected trade relationships (Section~3.2.1).} Section~\ref{sec:bipartivity} defines spectral bipartivity, a measure that quantifies the over-representation of even paths in these networks. Functional networks have high values of this measure compared to random expectation, as explained and defined in Section~\ref{sec:randomization}. In Section~\ref{sec:stats} we detail the statistical tools that we use to conduct this comparison. This methodology tested in Section~\ref{sec:implementation} on two public network datasets, producing the expected outcome.

\replaced{}{\text{\helvetica 3.2.1 Trade relationships}}

\subsubsection{Spectral bipartivity} \label{sec:bipartivity}

Spectral methods can be used to summarize the local connectivity structure of a network. The Estrada index is an absolute measure of local connectivity. This measure quantifies the local density of cycles by having closed paths contribute progressively less to the value of the measure, as they take more steps to complete~\citep{estrada_characterization_2000}. The value of the Estrada index for a network, $G$ with $n$ nodes can be computed as the trace of the matrix exponential of that network's adjacency matrix, $A$. Equation~\ref{eqn:Estrada} gives this definition as well as an alternative formulation where $\lambda_{1} \leq \cdots \leq \lambda_{n}$ are the eigenvalues of $A$.

\begin{equation}
EE(G) 
= tr \exp(A)
= \sum_{j=1}^{n} e^{\lambda_j} 
\label{eqn:Estrada}
\end{equation}

Most relevant to functional networks is a variation of the Estrada index that separates the contribution of even and odd closed paths: spectral bipartivity ($b_s$). This is done using the hyperbolic sine and cosine matrix functions, which add up to the matrix exponential, as applied to a network's adjacency matrix. With proper normalization, spectral bipartivity ranges from $0$ when the network is fully complete to $1$ when the network is fully bipartite~\citep{estrada_spectral_2005,estrada_protein_2006,estrada_network_2016}. Two-mode networks are the extreme case where the bipartite constraint on link formation entirely disallows odd cycles. Equation~\ref{eqn:bipartivity} defines several equivalent formulations of spectral bipartivity. 

\begin{equation}
b_s(G)
= \frac{tr \cosh(A) - tr \sinh(A)}{tr \cosh(A) + tr \sinh(A)} 
= \frac{tr \exp(-A)}{tr \exp(A)}
= \frac{\sum_{j=1}^{n} e^{-\lambda_j}}{\sum_{j=1}^{n} e^{\lambda_j}} \label{eqn:bipartivity}
\end{equation}

We primarily consider the value of the spectral bipartivity under a logistic transformation, because this metric is restricted in range from $0$ to $1$ and the Estrada index in the denominator can become quite large. Indeed, it grows exponentially with the square root of the number of edges in certain cases~\citep{pena_estimating_2007}. 

Moreover, the logit spectral bipartivity can be readily approximated for networks that fulfill the following two conditions. First, when the spectral bipartivity is very small, i.e. $b_s \approx 0$, the logistic transformation is closely approximated by a log transformation. Second, whenever the most positive and most negative eigenvalues of a network's adjacency matrix are substantially larger in magnitude than their neighboring eigenvalue, i.e.\ $\lambda_1 \ll \lambda_2$ and $\lambda_{n} \gg \lambda_{n-1}$, these eigenvalues will dominate the exponential sums in the numerator and denominator of spectral bipartivity. Equation~\ref{eqn:approx} describes these approximations.

\begin{equation}
\everymath{\displaystyle}
\begin{array}{c}
\textrm{logit}(b_s) 
 =  \log{\frac{b_s}{1-b_s}}
= \log{b_s} - \log{(1-b_s)}
\approx \log{b_s}\\
\log(b_s)  =  \log{\frac{\sum_{j=1}^{n} e^{-\lambda_j}}{\sum_{j=1}^{n} e^{\lambda_j}}} 
\approx \log{\frac{e^{-\lambda_1}}{e^{\lambda_n}}}
= - (\lambda_1 + \lambda_n)
\end{array}
\label{eqn:approx}
\end{equation}

\subsubsection{Identifying functional structure}
\label{sec:randomization}

\begin{figure}[h!] 
    \begin{center}
    \includegraphics[width=\textwidth]{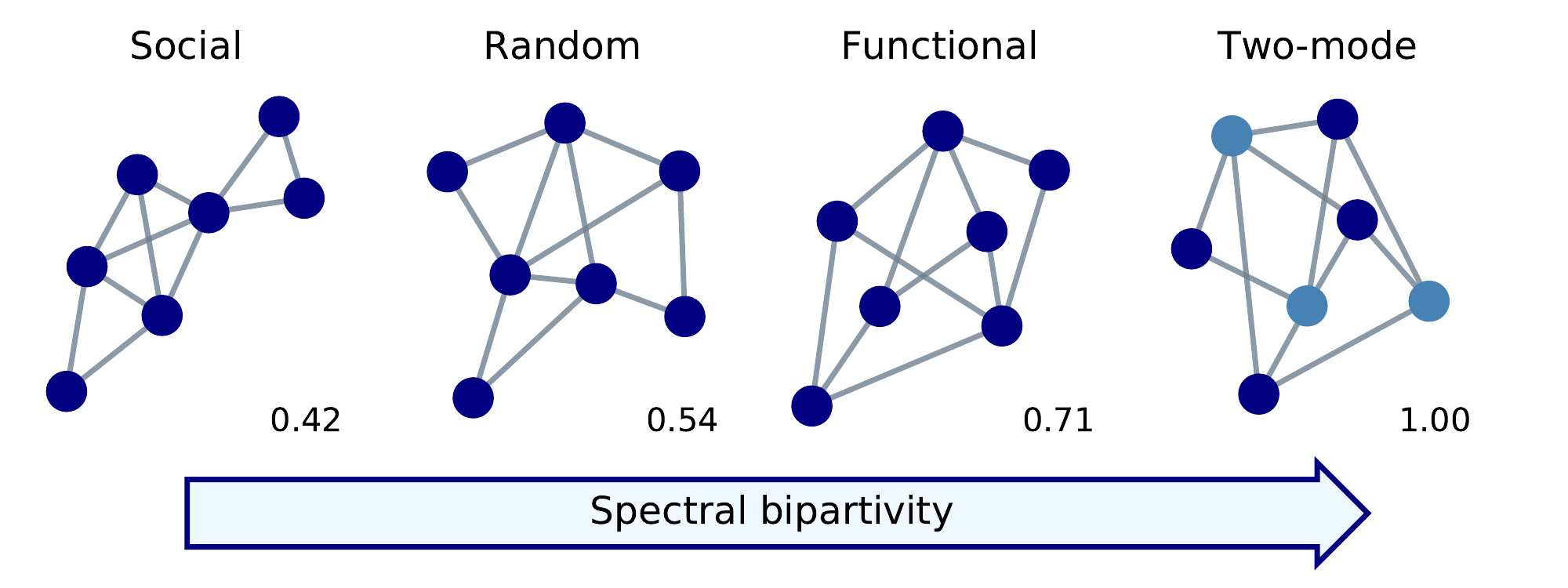}
    \end{center}
    \caption{Toy networks with seven nodes and eleven edges, each of a different type, shown in increasing order of their spectral bipartivity value.}
    \label{fig:scale}
\end{figure}

Here we propose that spectral bipartivity can be re-purposed to identify functional structure in networks using a comparison to random expectation. The measure quantifies the over-representation of even paths in the local connectivity structure of a network. Recall that functional networks have especially many squares with even path length, while social networks have especially many triangles with odd path length (see Section~\ref{sec:typology}). Random networks, with neither social nor functional structure, would produce values of spectral bipartivity that fall in-between those of the other two network types. Figure~\ref{fig:scale} gives a toy example of how social, random, functional, and two-mode networks with the same number of nodes and edges are arranged according to their value of spectral bipartivity. Notably, functional networks are more bipartite than random expectation. 

The random expectation is found by calculating spectral bipartivity on a set of random networks comparable to our networks of inferred trading relationships. Degree-preserving randomization~\citep{rao_markov_1996,milo_uniform_2004,gionis_assessing_2007} produces random networks that maintain the number of companies, the number of unique inter-company links, and the degree of each company. We use a version of this called random pairwise rewiring, wherein pairs of edges are selected and an end point of each edge are swapped~\citep[see:][Figure 1(a)]{hanhijarvi_randomization_2009}. In Section~\ref{sec:ensemble}, edge swaps were used to remove self-loops. Here, edge swaps randomize the network. Our implementation guarantees that the randomized network remains simple by following through with an edge swap only so long as it will not introduce self-loops or multi-edges. Randomization continues until $10 \cdot m$ pairs of links have been rewired, where $m$ is the number of simple edges. 

\subsubsection{Statistical test} \label{sec:stats}

We use non-parametric statistics to confirm that the networks described in Section~\ref{sec:data} have functional structure. Section~\ref{sec:bipartivity} defines spectral bipartivity and Section~\ref{sec:randomization} describes how we generate randomized versions of our networks. The spectral bipartivity computed on these networks are samples drawn from the reference distribution of this measure, as in a Monte Carlo test~\citep[see:][discussion by Barnard, G. A. pg.~294]{besag_generalized_1989,besag_sequential_1991,bartlett_spectral_1963}. For statistical comparison, we use the two-sample one-tailed Kolmogorov-Smirnov (KS) test~\citep{smirnov_table_1948,noauthor_kolmogorovsmirnov_2008,virtanen_scipy_2020}. 

The KS statistic quantifies the (lack of) overlap between distributions; $KS = 0$ indicates identical distributions and $KS = 1$ indicates non-overlapping distributions, i.e. one is consistently larger than the other. The statistical power of the test is a function of the size of the two distributions and the overlap between then. In analyzing the reconstructed networks, we compare their values of spectral bipartivity against those of their many randomized versions. Finding values consistently and significantly larger-than-random would indicate functional structure. In analyzing our multi-layer complementary configuration models, we consider the difference in logit-transformed spectral bipartivity between each instance and a randomized version of itself. Finding this difference to be consistently and significantly greater than zero would indicate that the model introduces functional structure. Finding one model's distribution to be consistently and significantly larger than another would indicate that it introduces functional structure to a greater extent. Note that comparisons between two distributions have much higher statistical power under non-parametric tests than those between a distribution and a single value.

\subsubsection{ \added{Implementation}} \label{sec:implementation}

\begin{figure}[h!] 
\begin{center}
\includegraphics[height=5.25cm]{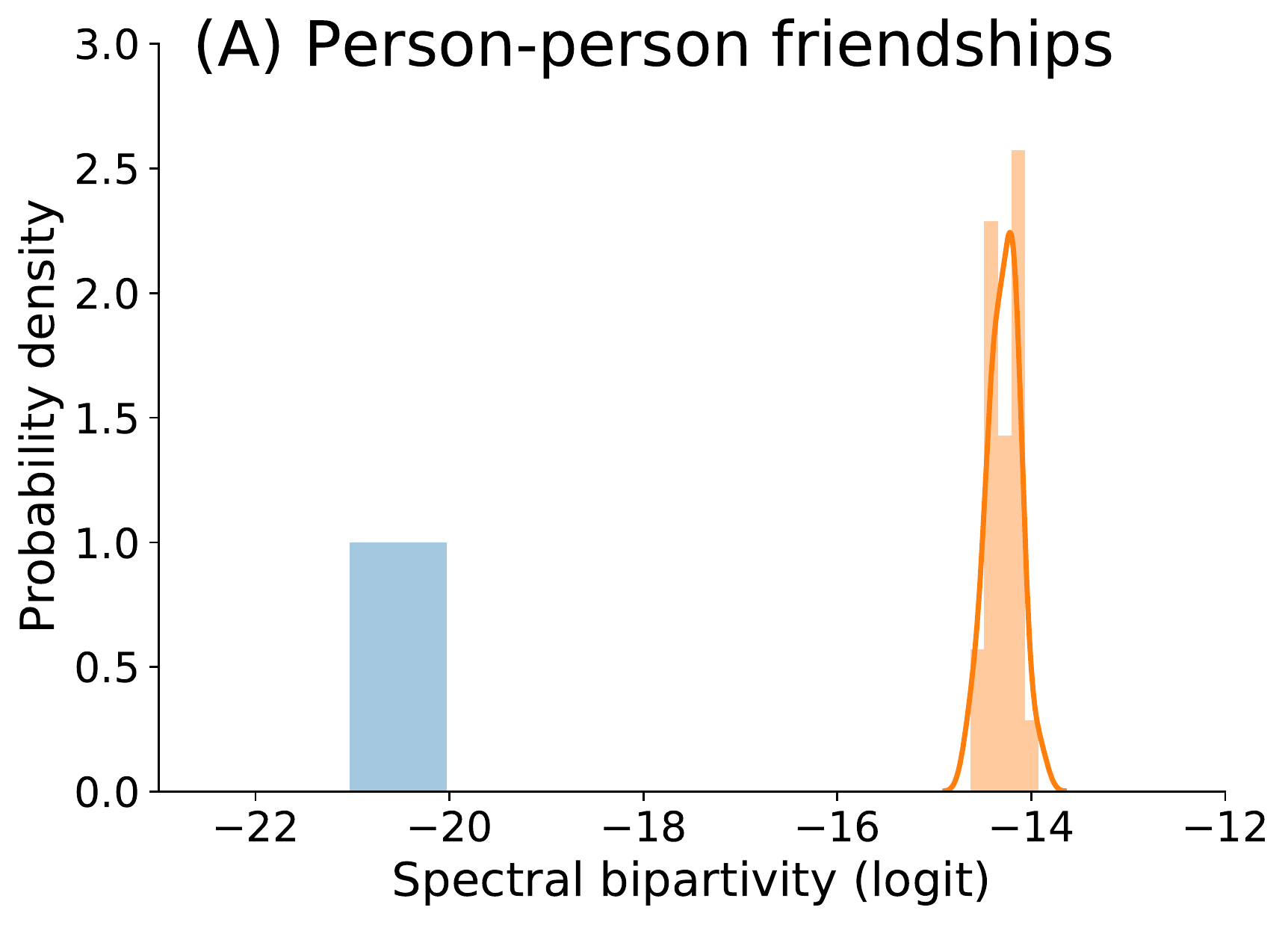}
\includegraphics[height=5.25cm]{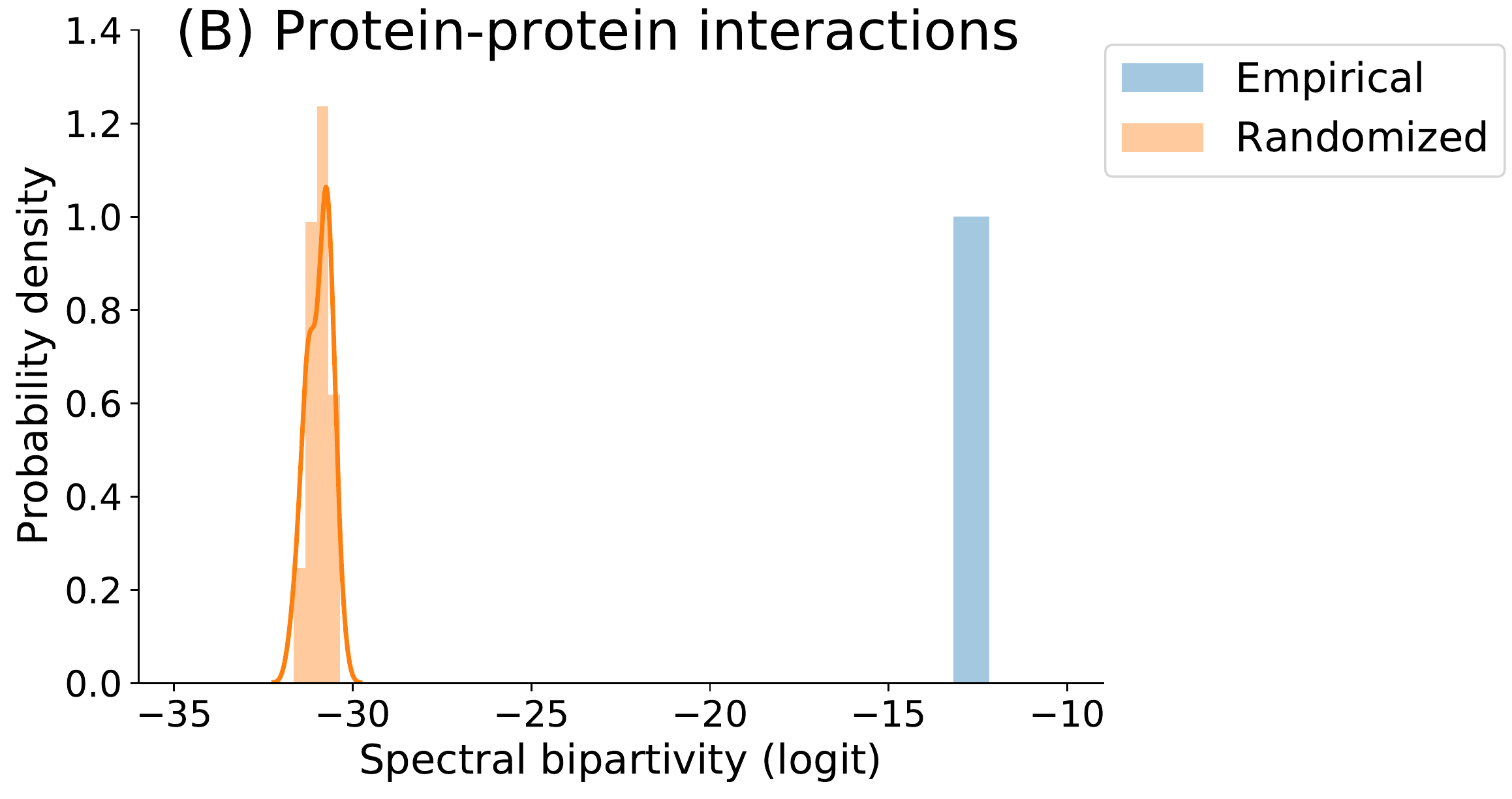}
\end{center}
\caption{\added{Value of logit-transformed spectral bipartivity for networks of (A) Person-person friendships and (B) Protein-protein interactions and the comparable distributions of their randomized versions.}}\label{fig:demo}
\end{figure}

\added{Our implementation of degree-preserving randomization, spectral bipartivity, and its (logit) approximation use \texttt{networkx}~\citep{hagberg_exploring_2008} and \texttt{scipy}~\citep{virtanen_scipy_2020}\replaced{.}{, enabling proper representation of the potentially very small values.} The methodology here described produces the expected result in a demonstration on two public network datasets. The first is a network of Facebook friendships among a group of first-year university students, collected as a part of the Copenhagen Networks Study~\citep{sapiezynski_interaction_2019}. Figure~\ref{fig:demo}(A) shows its spectral bipartivity is less than random expectation, indicating \emph{social} structure. The second is a network of interactions among human proteins, which was analyzed in \citet{kovacs_network-based_2019} and is available in that paper's supplementary material. Figure~\ref{fig:demo}(B) shows its spectral bipartivity is greater than random expectation, indicating \emph{functional} structure. The code to produce these figures is made available at \url{https://github.com/carolinamattsson/local-connectivity-structure}.}

\section{Results}
\label{sec:results}

In this section we present our analysis of local connectivity structure in company-level production networks. The reconstructed Zeeland and Netherlands networks show functional structure  (Section~\ref{sec:results:structure}) and this arises, at least in part, due to the complementary nature of customer-supplier ties (Section~\ref{sec:results:complementarity}).

\subsection{Local connectivity structure}
\label{sec:results:structure}

\begin{figure}[h!] 
\begin{center}
\includegraphics[height=5.25cm]{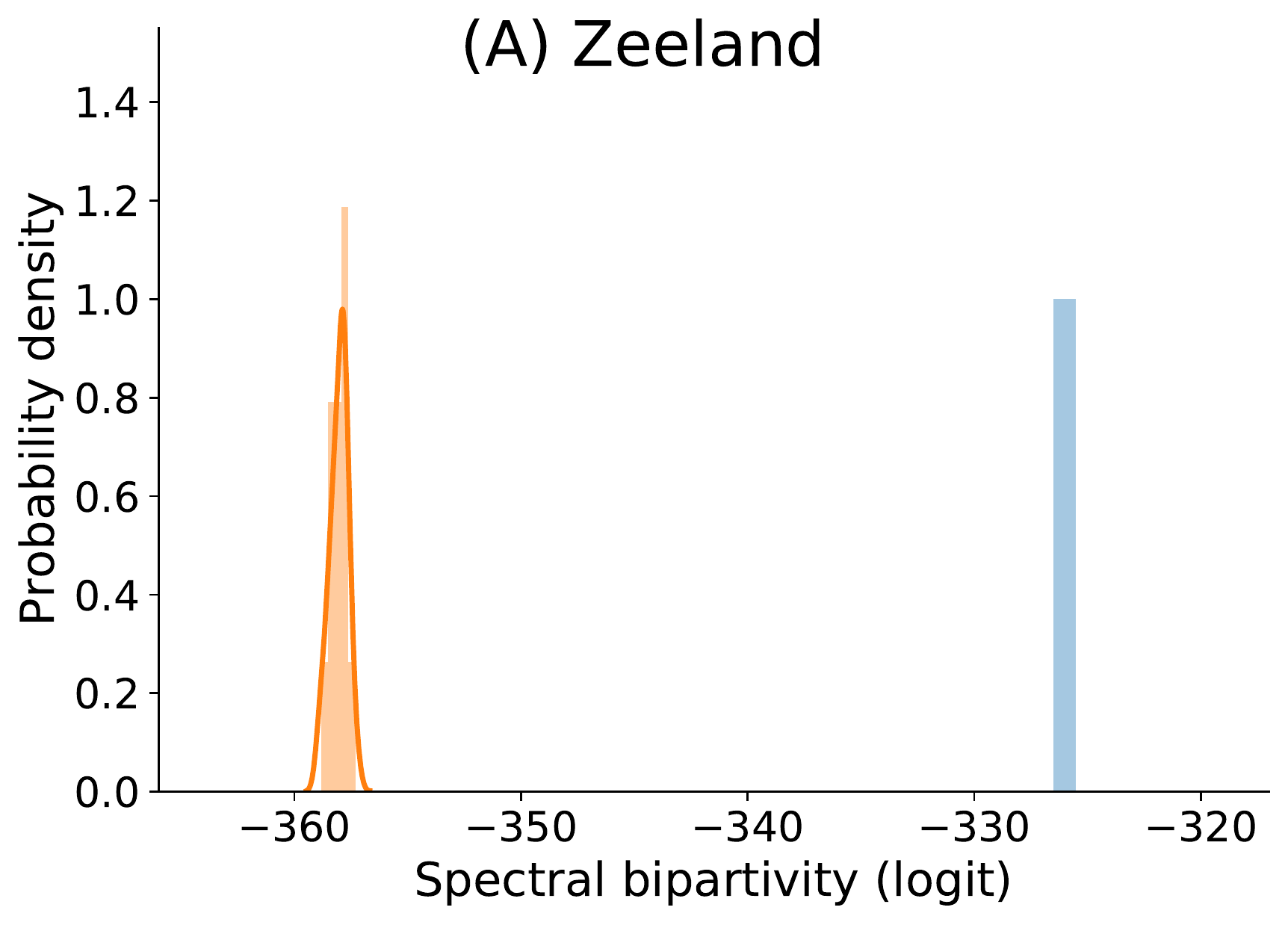}
\includegraphics[height=5.25cm]{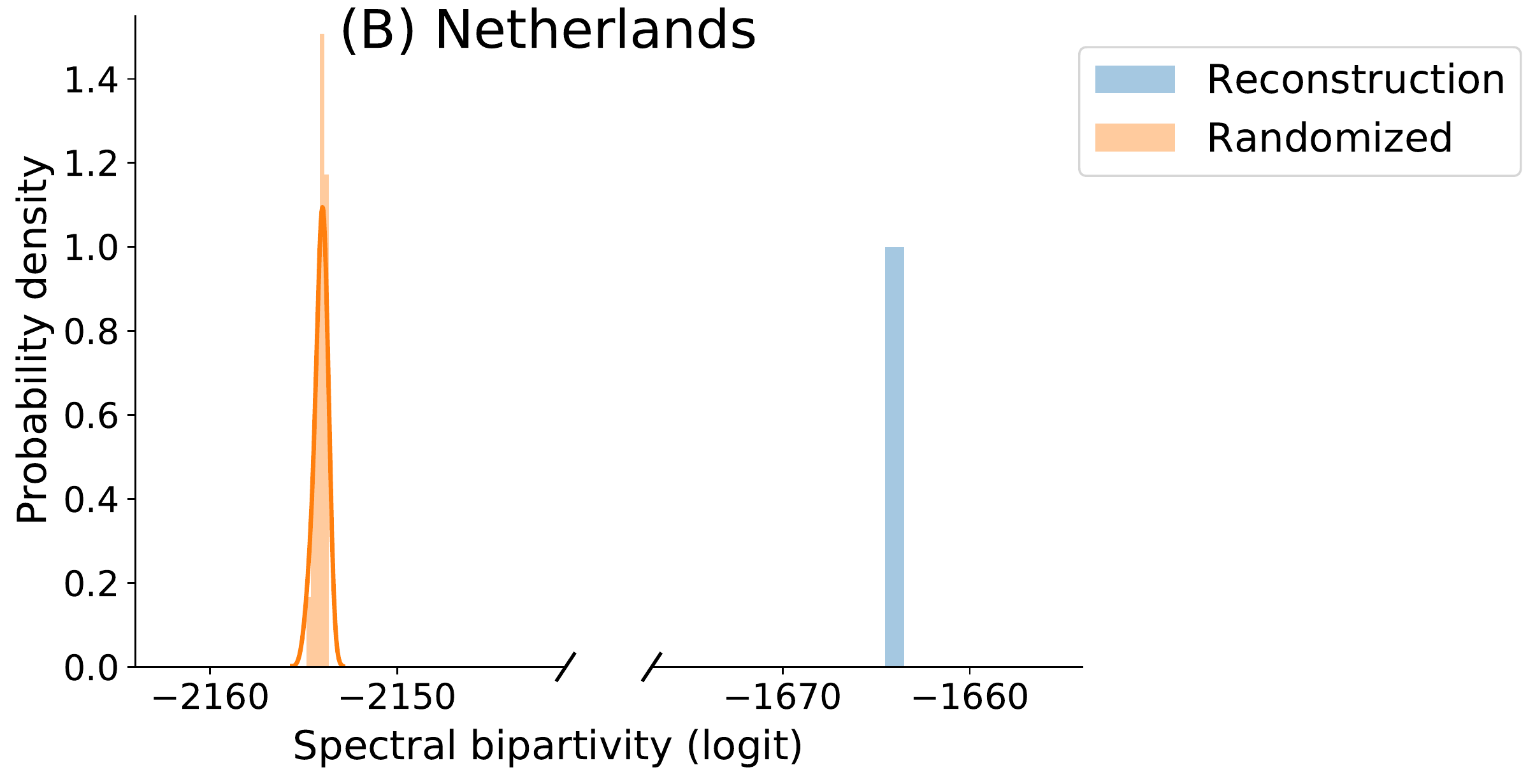}
\end{center}
\caption{Value of logit-transformed spectral bipartivity of reconstructed (A) Zeeland and (B) Netherlands production networks and the comparable distributions of their randomized versions.}\label{fig:bipartivity}
\end{figure}

We find functional structure in the reconstructed company-level production networks. The spectral bipartivity of the network of \added{(inferred)} trade relationships in Zeeland is substantially larger than random expectation; statistical comparison yields a one-tailed Kolmogorov-Smirnov statistic of $1.0$ ($N_1 = 1$, $N_2 = 1000$, $p < 0.001$). Figure~\ref{fig:bipartivity}(A) plots the logit-transformed value of spectral bipartivity for the Zeeland production network against the distribution of values over $25$ randomized instances. The network falls well above random expectation and can be said to have functional structure. The value of spectral bipartivity (in the absolute sense) is small, at $1.6 \cdot 10^{-142}$, indicating a lack of bipartite structure and that the network is definitely not two-mode.

Using the approximation described in Equation~\ref{eqn:approx}, we extend this result to the company-level production network for the whole of the Netherlands (Figure~\ref{fig:bipartivity}(B)). This network, consisting of 50,930,077 inferred trade relationships among 102,461 companies with $5+$ employees, also has functional structure. Its logit spectral bipartivity is significantly larger-than-random ($KS = 1.0$, $N_1 = 1$, $N_2 = 25$, $p < 0.04$). 

\added{In interpreting these results, recall that the comparison to random expectation is key to our re-purposing of spectral bipartivity for analyzing local connectivity structure. Values of spectral bipartivity, in the absolute sense, reflect also other network-structural features. In particular, this measure is defined as a ratio and the denominator of this ratio can be strongly affected by the number of edges in the network (see~Section~\ref{sec:bipartivity}). The Zeeland network has a higher value of spectral bipartivity than the Netherlands network because the Netherlands network is substantially larger in size. Randomization preserves the number of edges (and the degree sequence) so it is serving as a way to ``center'' the scale such that remaining differences reflect local connectivity structure; this lets us identify functional structure. Developing measures that would allow for comparison across networks substantially different in size is a promising area for future work.}

\replaced{}{Identifying functional structure in these reconstructed networks offers a new light in which to consider two existing findings from previously published papers on empirical company-level production networks. \citet{ohnishi_network_2010} study three-node motifs in a Japanese customer-supplier network from 2005 provided by Tokyo Shoko Research Ltd. Compared to random expectation, they find two-link V-shaped motifs to be over-represented while three-link feedforward and feedback loops are substantially and significantly under-represented. This result implies fewer-than-expected triangles in the local connectivity structure of the empirical network, which we interpret as evidence of functional structure. \citet{fujiwara_large-scale_2010} show disassortativity in degree for a similar network from the same data supplier. This result lends credence to the idea that disassortativity is an expected feature of functional networks, more broadly.}

\subsection{Customer-supplier complementarity}
\label{sec:results:complementarity}

We find evidence that functional structure in production networks is driven by 
\added{the principle of node complementarity, that is,} the practical fact that trade  \replaced{relationships imply the exchange of some product from a supplier to a customer}{involves the supply and use of particular products}. Our multi-layer configuration models allow customer-supplier ties only between pairs of companies compatible in trade; a likely supplier and a likely user of products with a particular classification (see~Section~\ref{sec:ensemble}). Each of the resulting networks reflect trade-compatible relationships between companies with complementary roles in this network. This constraint gets progressively more stringent as the product classification that defines the roles gets more detailed (see Table~\ref{tab:model}).

\begin{figure}[!htb] 
    \begin{center}
    \includegraphics[height=5.25cm]{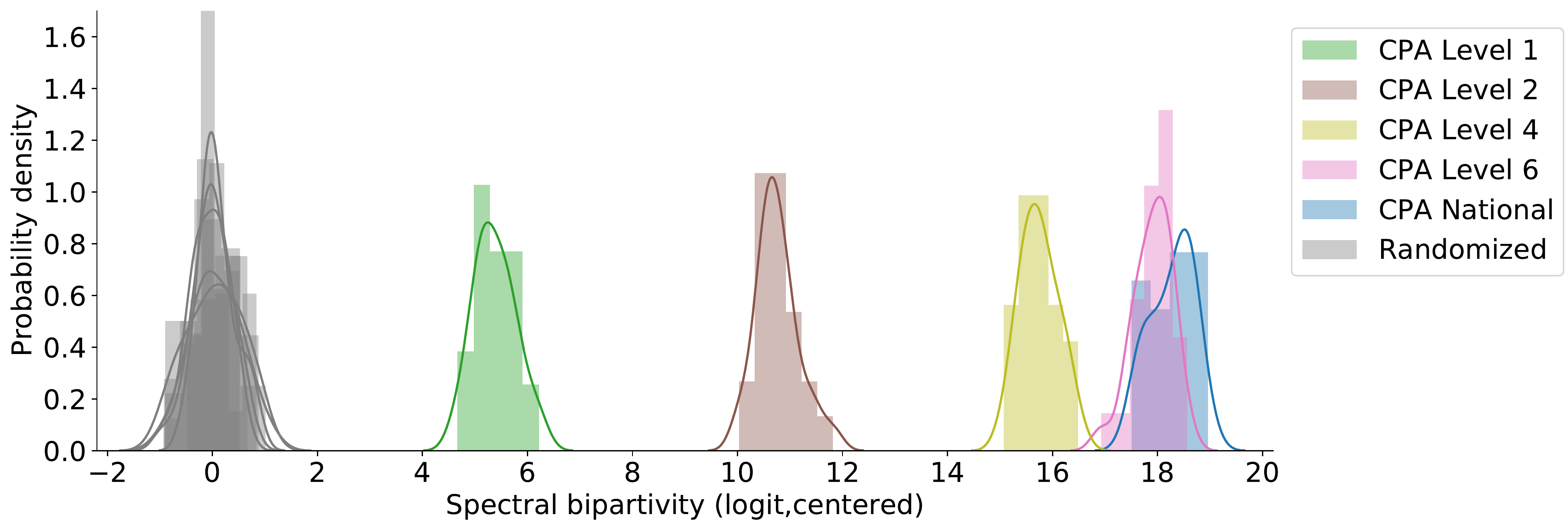}
    \end{center}
    \caption{Values of logit-transformed spectral bipartivity for simulated networks of trade-compatible relationships and their randomized comparisons, centered by the median comparison value. Layers are designated according to the European CPA (2008) and its Dutch implementation, which is the most detailed.}
    \label{fig:complementarity}
\end{figure}

Already at the top, sector-level, classification (CPA Level 1) \replaced{customer-supplier}{the} complementarity \replaced{}{constraint} introduces some functional structure. Figure~\ref{fig:complementarity} compares $25$ instances from our configuration model at CPA Level 1 to their randomized versions, where the relationships are no longer necessarily trade-compatible. The logit-transformed spectral bipartivity is consistently larger-than-random where \added{customer-supplier} complementarity has been introduced at CPA Level 1 ($KS = 1.0$ , $N_1 = 1$, $N_2 = 25$, $p < 0.04$). Stricter complementarity constraints at the more detailed CPA Levels 2, 4, and 6 each introduce consistently and significantly more functional structure ($KS = 1.0$ , $N_1 = 25$, $N_2 = 25$, $p < 10^{-14}$). On the other hand, and as expected, the small refinement between CPA Level 6 and the CPA National product classification produces a minor increase in the separation between model and random expectation; it is statistically, but not substantively, significant ($KS = 0.44$ , $N_1 = 25$, $N_2 = 25$, $p < 0.01$).

\added{In interpreting this finding, recall that these are simulated networks whose links are sampled from among the many possible links between trade-compatible companies. The roles of individual companies are defined with respect to standardized industry and product categorizations used in the collection of official economic statistics throughout Europe~\citep{eurostat_nace_2008}. These categories are broad relative to the number of differentiated products traded within the Netherlands and so trade relationships in our networks reflect potential trade in a number of relevant products. This means that the constraint we impose in generating our networks is fairly blunt, even at CPA National and especially at CPA Level 1. In this way, our generalization of the CBS network reconstruction process serves as an important robustness check on the results in Section~\ref{sec:results:structure}. At the same time, functional structure could come about via various channels. Most trivially, two suppliers and two users of the same product could form a closed square due to redundant purchasing patterns. This is unlikely to be what we are picking up on as it is assumed that the vast majority of companies deal with only one or a few suppliers per product-group (see Section~\ref{sec:networkconstruction}). The commonest situation forming a closed square, here, would be one where two suppliers of two different products sell those products to the same two customers. This interpretation suggests a potentially fruitful line of inquiry relating the principle of complementarity on networks as defined in \citet{kitsak_latent_2020} with the notion of product complementarity as defined in the economics literature.}

\section{Discussion} \label{sec:discussion}

This paper has \added{advanced the hypothesis that company-level production networks are so-called ``functional'' networks, with a distinctive local connectivity structure first identified in network biology~\citep{kovacs_network-based_2019,kitsak_latent_2020}. This hypothesis was made concrete through a discussion on local connectivity structures and the re-interpretation of previous empirical findings through this lens~\citep{ohnishi_network_2010,fujiwara_large-scale_2010}. We then explored this hypothesis, directly, using a re-purposed network metric~\citep{estrada_network_2016} and an existing dataset of inferred customer-supplier ties produced by Statistics Netherlands (CBS)~\citep{hooijmaaijers_methodology_2019}. This methodology identifies functional structure in production networks representing trade among companies in Zeeland province and the whole of the Netherlands. Our generalization of the CBS network reconstruction process, using multi-layer configuration models, then illustrates that customer-supplier complementarity is key to the emergence of functional structure in company-level production networks.}

\added{In interpreting these findings, our company-level production networks should be understood as a selection of likely trade relationships where the industry pairing implies the two companies are compatible in trade. Much nuance is avoided in that the industry and product categorizations used in the generation of these networks are standardized and broad; inferred ties reflect potential trade in a number of relevant products. However, as noted in Section~\ref{sec:networkconstruction}, inferences about individual companies are noisy as the underlying data are collected for statistical purposes. For this reason, we limit our network analysis to characterizing whole-network patterns in local connectivity structure. To illustrate functional structure in production networks at a more intuitive scale, Section~\ref{sec:literature} highlights prior work that finds closed squares and interpretable functional modules in empirical data; it is reassuring that this fully supports our conclusions. The assumptions made in the CBS network reconstruction process when matching customers with suppliers could more directly affect the local connectivity structure of our networks. On this point, it is reassuring that functional structure persists also under our generalized reconstruction process and using less detailed product classifications. Exploring a wider range of data-informed network generation processes would allow more specific hypotheses to be tested and is a promising direction for future work.}

\replaced{Our findings}{let us conclude, with some generality, that company-level production networks have functional structure.} have practical implications for the analysis of company-level production networks and wider implications for the study of economic systems. Practically speaking, our work helps narrow down the set of network analysis techniques relevant for use with networks representing trade relationships among companies, be they empirical, reconstructed, or modeled. \added{To be interpretable, these should conform to the logic of functional networks:} companies trade with complementary others and it is close competitors---companies with many shared customers and suppliers---who are especially similar. For the problem of link prediction, for instance, heuristics that close squares (L3) are expected to be especially accurate~\citep{kovacs_network-based_2019}. On the other hand, there is little reason to expect high levels of clustering or high levels of reciprocity between customers and suppliers. Intuitions and techniques developed within network biology for use with PPI networks are likely to be especially applicable~\citep{barabasi_network_2004,kitsak_latent_2020}. For instance, identifiable meso-scale structures can be interpreted more readily as ``functional modules''~\citep{chen_detecting_2006} than as ``communities''~\citep{blondel_fast_2008}. \replaced{}{The study of company-level production networks stands to benefit from advances in network biology.}

\replaced{}{With respect to network science, we demonstrate the usefulness of network categorization according to local connectivity structure. The intuitions and techniques developed within network biology are applicable to a wider set of domains that study functional networks~\citep{barabasi_network_2004,kitsak_latent_2020}. Companies trade with complementary others and it is close competitors---companies with many shared customers and suppliers---who are especially similar. Relating several application areas where networks are shaped by this same principle might advance our understanding of how local connectivity structure affects also meso-scale and whole-network structure. Doing so will involve developing better measures and tools tailored especially for functional networks.}

\replaced{Speaking more broadly}{With respect to complexity economics}, functional structure can deepen our understanding of dynamics on and of production networks. The short-term dynamics of ways economic systems fail are already studied using detailed simulations over empirical company-level production networks~\citep{hazama_measuring_2017,inoue_propagation_2020}. Supply disruptions are known to compound when also the competitors of affected companies become affected via shared customers and suppliers, a phenomenon not unrelated to the prominence of closed squares in the local connectivity structure. An anecdotal example is that during the financial crisis in late 2008 the CEO of Ford, Alan Mulally, gave testimony in favor of US Government support for General Motors and Chrysler. Mulally argued that the demise of his competitors would imperil shared suppliers of highly specialized auto parts~\citep[][pg.~145]{klier_restructuring_2013}. \added{Note that the Japanese automotive industry, also, is especially locally bipartite~\citep[pg.~570]{fujiwara_large-scale_2010}.} In the demand direction, \added{one might consider the analogous phenomenon where} Chinese e-commerce platforms selling personal protective equipment to many of the same customers all experienced shortages in sourcing from many of the same manufacturers following reports of a deadly outbreak of infectious disease in Wuhan, China in January, 2020~\citep{mcmorrow_chinese_2020}. 

\added{Our updated understanding of the local connectivity structure of production networks} can help us reason about the \replaced{}{microeconomic} mechanisms behind economic dynamics also over the medium- and long- term. \added{Within the economics literature, studies of micro-economic mechanisms and macro-economic dynamics remain largely separate.} Periodic fluctuations in aggregate output and economic growth over time are thought to be driven by productivity improvement at the level of industries~\citep{acemoglu_network_2012,carvalho_production_2018}. \added{With an understanding of functional structure, one might consider the impact of closed squares on the establishment of} price information and trust, both of which are required for companies to realize productivity improvements~\citep{uzzi_social_1997,cardoso_trading_2019}. Likewise, much is known about the evolutionary dynamics of economic systems at the industry level~\citep{hausmann_atlas_2013,mcnerney_how_2018,mealy_economic_2020,dam_variety_2020} and from case studies~\citep{coenen_local_2010,dawley_creating_2014,boschma_towards_2017}. \added{In bridging these scales it will be key to consider the emergence of meso-scale structures that perform higher-level functions within production networks.} \added{Finally, functional structure in production networks opens doors to insights from previously unrelated domains also studying the routine function, failure, and growth of functional networks.}

\section*{Conflict of Interest Statement}
The authors declare that the research was conducted in the absence of any commercial or financial relationships that could be construed as a potential conflict of interest.

\section*{Author Contributions}
EH, CM, \& FT developed the theory. CD, CM, \& FT determined the method. GB \& CM performed the analysis. AF \& PS managed the project. CM wrote the manuscript. All authors discussed the results and edited the final manuscript.

\section*{Funding}
This work was supported by the Ministry for Economic Affairs and Climate, The Netherlands.

\section*{Acknowledgments}
The authors acknowledge Joyce Ten Holter and Gideon Mooijen for valuable assistance. We thank Leo Torres, István Kovács, and Maksim Kitsak for discussion.


\section*{Data Availability Statement}
The data analyzed for this study is under the administration of Statistics Netherlands (CBS). Requests to access these datasets should be directed to Gert Buiten, g.buiten@cbs.nl.

\bibliographystyle{frontiersinSCNS_ENG_HUMS} 
\bibliography{references}

\end{document}